\documentclass[a4paper,fleqn,usenatbib]{mnras}
\usepackage[T1]{fontenc}
\usepackage{ae,aecompl}

\usepackage{graphicx}	
\usepackage{amsmath}	    
\usepackage{amssymb}	    

\title[Exotic Image Formation in Cluster Lenses]{Exotic Image Formation in Strong Gravitational Lensing by Clusters of Galaxies. I: Cross-Section}

\author[A. K. Meena and J. S. Bagla]{
Ashish Kumar Meena,$^{1}$\thanks{E-mail: ashishmeena766@gmail.com}
Jasjeet Singh Bagla,$^{1}$\thanks{E-mail: jasjeet@iisermohali.ac.in}
\\
$^{1}$Indian Institute of Science Education and Research Mohali,
Knowledge City, Sector 81, Sahibzada Ajit Singh Nagar, Punjab 140306,
India\\}

\pubyear{2021}

\begin{document}
\label{firstpage}
\pagerange{\pageref{firstpage}--\pageref{lastpage}}
\maketitle

\begin{abstract}
In a recent paper we have discussed the higher order singularities in
gravitational lensing.
We have shown that a singularity map, comprising of $A_3$-lines and
unstable (point) singularities ($A_4$ and $D_4$), is a compact
representation of high magnification regions corresponding to a given
lens model for all possible source redshifts.  
It marks all the optimal locations for deep surveys in the lens
plane. 
Here we present singularity maps for ten different clusters
lenses selected from the \textit{Hubble Frontier fields} (HFF) and the
\textit{Reionization Lensing Cluster Survey} (RELICS) surveys.
We have identified regions in the lens plane with a high magnification
for sources up to redshift ten. 
To determine the dependence of unstable (point) singularities on lens
mass model reconstruction techniques, we compared singularity maps
corresponding to the different mass models (provided by various groups
in the HFF survey) for each cluster lens.  
We find that the non-parametric (free-form) method of lens mass
reconstruction yields the least number of point singularities.  
In contrast, mass models reconstructed by various groups using a
parametric approach have a significantly larger number of point
singularities. 
We also estimate the number of galaxies lying near these unstable
(point) singularities, which can be observed with the \textit{James
  Webb Space Telescope} (JWST). 
We find that we expect to get at least one hyperbolic umbilic and one
swallowtail image formation for a source at $z > 1$ for every five
clusters with JWST.
These numbers are much higher than earlier estimates. 
\end{abstract}

\begin{keywords}
gravitational lensing: strong -- galaxies: clusters: individual (Abell 370, 
Abell 2744, Abell S1063, MACS J0416.1-2403, MACS J1149.5+2223, MACS J0159.8-0849, MACS J0308.9+2645, PLCK G171.9-40.7, PLCK G287.0+32.9, SPT-CLJ0615-5746)
\end{keywords}

\section{Introduction}
\label{sec:Introduction}

Galaxy clusters with strong gravitational lensing are a very powerful
tool to study the physics of the Universe~\citep{1992ARA&A..30..311B}.  
The large magnification factor in cluster lenses allows us
to observe distant sources (which otherwise would have remained
unobserved) and helps us understand the evolution of galaxies in
the universe \citep{2010Sci...329..924J, 2011A&ARv..19...47K,
  2018MNRAS.479.5184A}.   
On the other hand, the strongly lensed background objects help us to
constrain the mass distribution within these cluster lenses, hence,
helping us to understand different processes going on inside these
galaxy clusters~\citep{2005A&A...442..405P, 2019A&A...631A.130B}.  
After the discovery of the first cluster lens
system~\citep{1988A&A...191L..19S}, extensive work has been done
dealing with both theoretical and observational aspects of cluster
lensing.  
On the theoretical side, different methods have been developed to
reconstruct the cluster mass distribution using the strongly lensed
background sources~\citep{2005ApJ...621...53B, 2005MNRAS.360..477D,
  2007NJPh....9..447J, 2007MNRAS.380.1729L, 2008ApJ...681..814C},  
and detailed studies have been done to use cluster lenses to probe 
the Universe~\citep{1998A&A...330....1B, 2013SSRv..177...31M,
  2016MNRAS.457.2738B}.  
Several surveys have been carried out to increase the number of known
strong lens systems in order to improve the quality of theoretical
findings~(CASTLES: \cite{1998Ap&SS.263...51M}, CLASS:
\cite{2003MNRAS.346..746C}, CLASH: \cite{2012ApJS..199...25P}, HFF:
\cite{2017ApJ...837...97L}, RELICS: \cite{2019ApJ...884...85C}). 

Almost all clusters behave as gravitational lenses. 
However, not every source lying behind a cluster is magnified by a 
large factor. 
Only sources lying near the caustics in the source plane are 
highly magnified. 
The factor by which a source is magnified also depends on the source
size: smaller the source, higher the magnification
factor~\citep{2018NatAs...2..334K}.  
For a given lens system, the caustic structure in the source plane is
sensitive to the source redshift. 
As a result, the area in the source plane that is highly magnified, 
changes as one varies the source redshift.
Unlike galaxies, clusters are very complex structures due to the
presence of numerous substructures. 
A galaxy can be modeled by using a single elliptical profile, whereas the 
modeling of cluster lenses needs to take into account the presence of 
various substructures like individual galaxies or groups of galaxies.  
This can also be seen in the evolution of caustic structure in the source 
plane with redshift. 
The evolution of caustics with redshift mainly includes the formation
and destruction of cusps, the exchange of cusps between radial and
tangential caustics, in such a way that the total number of cusps in
the source plane always remains even.   
Curves corresponding to caustics in the lens plane are known 
as critical curves. 
Highly magnified images of a strongly lensed source are formed in
close vicinity of these critical curves. 
These high magnification regions produced by cluster lenses can help us to 
observe galaxies at very high redshifts ($z>5$) including the first
galaxies.  
Such highly magnified systems have already been encountered in
different lensing surveys~\citep{2008ApJ...678..647B,
  2013ApJ...762...32C, 2015Natur.519..327W}.  
At present, the number of such systems is small as lensed galaxies are
very faint, and the number density of galaxies at these high redshifts
is small.  
However, the number of such strong lens systems is expected to
increase by more than an order of magnitude with the upcoming
facilities like EUCLID: \citep{2009arXiv0912.0914L}, JWST:
\citep{2006SSRv..123..485G}, LSST: \citep{2019ApJ...873..111I},
WFIRST: \citep{2019arXiv190205569A}.  

In our current work, we locate the highly magnified regions in the lens 
plane, for all source redshifts, for ten
different cluster lens systems from the \textit{Hubble Frontier Fields
  Survey} (HFF) and the \textit{Reionization Lensing Cluster Survey}
(RELICS).  
The algorithm to do this has been discussed briefly in 
\citet{2001ASPC..237...77B} and presented in detail in 
\citet{2020MNRAS.492.3294M} (hereafter paper I). 
As discussed in paper I, a singularity map consisting of $A_3$-lines
and unstable  (point) singularities is ideal for our study. 
These point singularities (swallowtail, hyperbolic umbilic, elliptic
umbilic) are formed only for some specific source redshifts and
specific source positions in the source plane.  
Apart from that, every point singularity comes with a characteristic
image formation.
$A_3$-lines correspond to cusp in the source plane and these are present
over a wide range of source redshifts. 
As cusps are stable singularities, these are continuous lines instead
of points in the singularity map, with points corresponding to
different source redshifts. 
The image formation corresponding to structures ($A_3$-lines and point 
singularities) in singularity maps (for the appropriate source redshift) 
shows three or more highly magnified images lying near each other in 
a small region of the lens plane in the vicinity of the singularity.
The singularity maps corresponding to cluster lenses not only point
out the highly magnified regions in the lens plane but are also
sensitive to the lens mass reconstruction techniques. 
Here we also compare different mass models corresponding to each
cluster lens.  
These mass models are reconstructed using different (parametric and 
non-parametric) methods.
The comparison has been done to see how sensitive the $A_3$-line
structure and the total number of point singularities is to the cluster
mass reconstruction method as different approaches use different sets
of underlying assumptions.  

Apart from locating the highly magnified regions in the lens plane and
looking at the effect of mass reconstruction methods on the
singularity map, the other important point that has been
discussed \citep{2009MNRAS.399....2O} is to estimate the expected
number of source galaxies lying near the point singularities.  
This allows us to estimate the probability of observing these
characteristic image formations in the upcoming large
scale surveys.  
In order to do so, we require the distribution of galaxies as a
function of redshift. 
This can be determined from the galaxy luminosity function (GLF) and
the Schechter function~\citep{1976ApJ...203..297S} is widely used to 
parametrize it.
Various studies using different surveys have been carried out to determine 
the rest frame GLF as a function of the redshift in different wavelength bands 
(UV:\cite{2018PASJ...70S..10O, 2020MNRAS.493.2059B, 2020MNRAS.tmp..668M}, 
IR: \cite{2007MNRAS.380..585C, 2010MNRAS.401.1166C, 2017MNRAS.465..672M}).
Different groups have estimated the number of galaxies that may be
observed with JWST considering different models of galaxy formation
and evolution~\citep{2018MNRAS.474.2352C, 2018ApJS..236...33W,
  2019MNRAS.483.2983Y}.  
Following~\cite{2018MNRAS.474.2352C}, we estimate the number of
exotic images that may be observed with JWST in one of the NIRCam
bands. 

This paper is organized as follows.
In \S\ref{sec:singularities}, we briefly review the basics of
the stable and unstable (point) singularities in gravitational
lensing.  
The cluster lenses used in the present analysis are enumerated in
\S\ref{sec:clsuter lenses}. 
The results are presented in \S\ref{sec:results}.
The construction of singularity maps for different cluster lenses is
discussed in \S\ref{ssec:singularity maps}.
Discussion of stability of singularity maps is presented in
\S\ref{ssce:stability}.  
In \S\ref{ssec:cross section}, we estimate the number of strongly
lensed galaxy sources with characteristic image formations near these
point singularities, that can be observed with the JWST.
In \S\ref{ssec: red_measure} we discuss the possibility of
constraining the source redshift using point singularities.
Summary and conclusions are presented in \S\ref{sec:conclusion}. 
We also discuss the future work in this section. 
The cosmological parameters used in this work to calculate the various 
quantities are: $H_0=70\:kms^{-1}Mpc^{-1},\:\Omega_\Lambda=0.7,\: 
\Omega_m=0.3.$

\begin{table*}
  \centering
  \caption{Cluster lenses used in current analysis: The upper half of
    the table lists the cluster lenses taken from the \textit{Hubble
      Frontier fields (HFF)} survey, whereas the lower part of the
    table has details of the cluster lenses from the
    \textit{Reionization Lensing Cluster Survey (RELICS)}.  
    For the HFF clusters, four different mass models provided by
    Keeton, Sharon, Williams, and Zitrin (zitrin$\_$nfw) groups are
    used.  
    For the RELICS clusters, we only use one mass model  for each cluster 
    provided by Zitrin group(zitrin$\_$ltm$\_$gauss). 
    The version and the resolution of these mass models is listed below.}
  \label{tab:cluster_lenses}
  \begin{tabular}{lccccr} 
    \hline
    \multicolumn{5}{|c|}{HFF Clusters} \\
    \hline
    & Keeton & Sharon & Williams & Zitrin \\
    \hline
    Abell 370 (A370)            & v4($0.06''$) & v4($0.05''$) & v4($0.05''$) & v1($0.050''$) \\
    Abell 2744 (A2744)          & v4($0.06''$) & v4($0.05''$) & v4($0.05''$) & v3($0.060''$) \\
    Abell S1063 (AS1063)        & v4($0.06''$) & v4($0.05''$) & v4($0.05''$) & v1($0.065''$) \\
    MACS J0416.1-2403 (MACS0416)& v4($0.06''$) & v4($0.05''$) & v4($0.05''$) & v3($0.060''$) \\
    MACS J1149.5+2223 (MACS1149)& v4($0.06''$) & v4($0.05''$) & v4($0.05''$) &  \\
    \hline
    \multicolumn{5}{|c|}{RELICS Clusters} \\
    \hline
    & Zitrin &  &  & \\
    \hline
    MACS J0159.8-0849 (MACS0159)  & v1($0.06''$) &  &  &  \\
    MACS J0308.9+2645 (MACS0308) & v1($0.06''$) &  &  &  \\
    PLCK G171.9-40.7 (PLCKG171   & v1($0.06''$) &  &  &  \\
    PLCK G287.0+32.9 (PLCKG287)  & v1($0.06''$) &  &  &  \\
    SPT-CLJ0615-5746 (SPT0615)   & v1($0.06''$) &  &  &  \\
    \hline
  \end{tabular}
\end{table*}

\begin{figure*}
  \includegraphics[width=\textwidth,height=8.5cm,width=8.5cm]{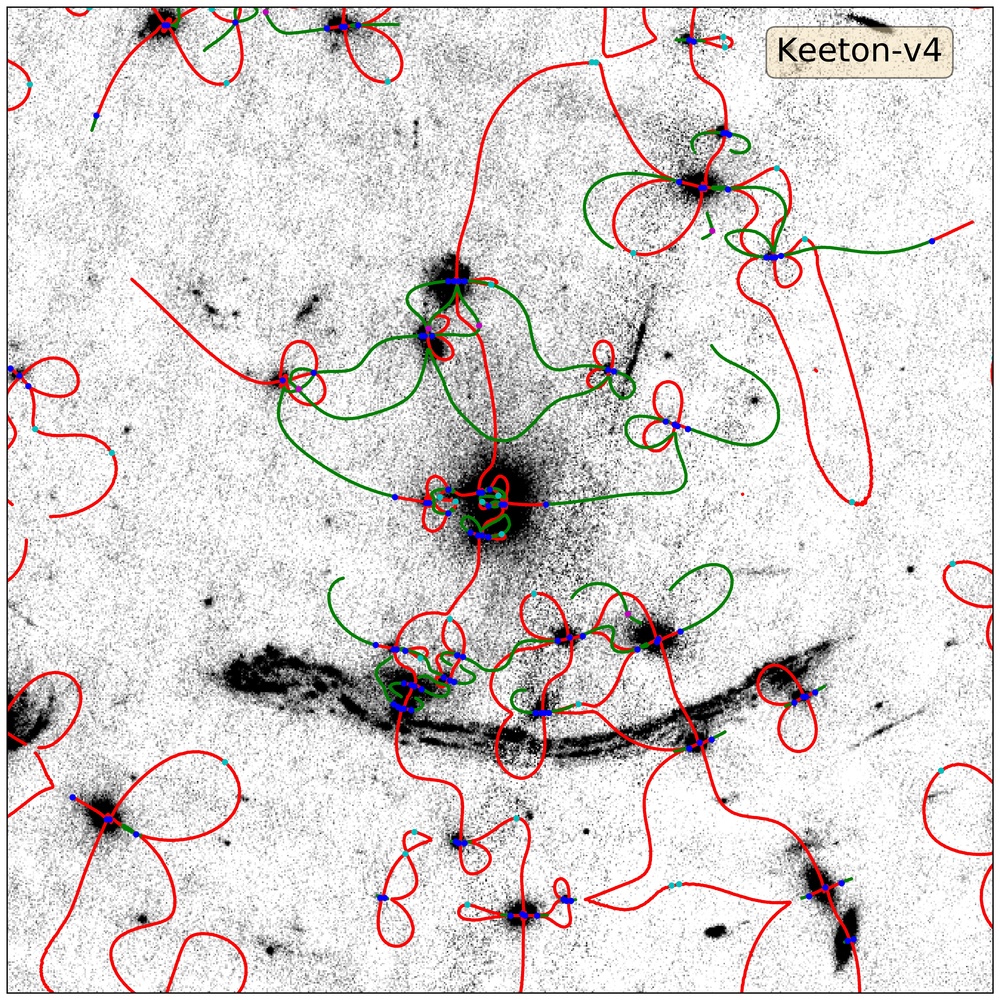}
  \includegraphics[width=\textwidth,height=8.5cm,width=8.5cm]{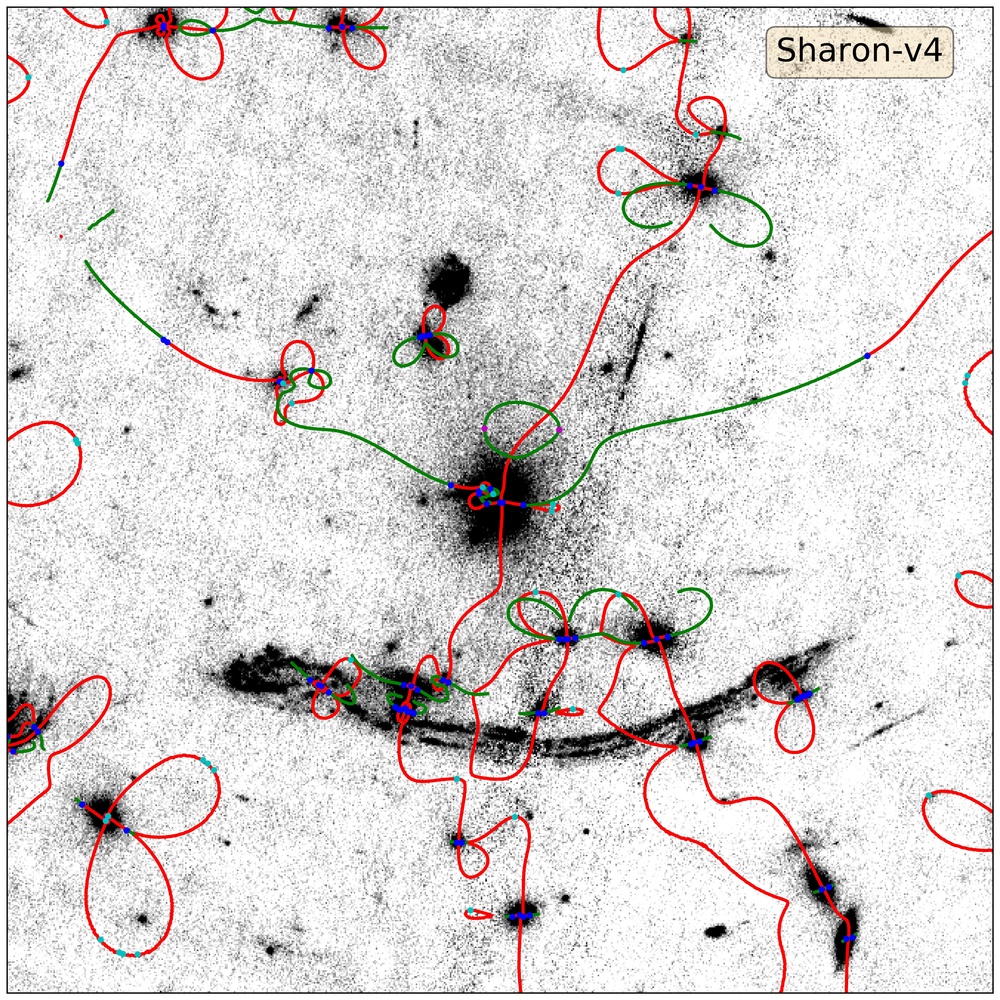}
  \includegraphics[width=\textwidth,height=8.5cm,width=8.5cm]{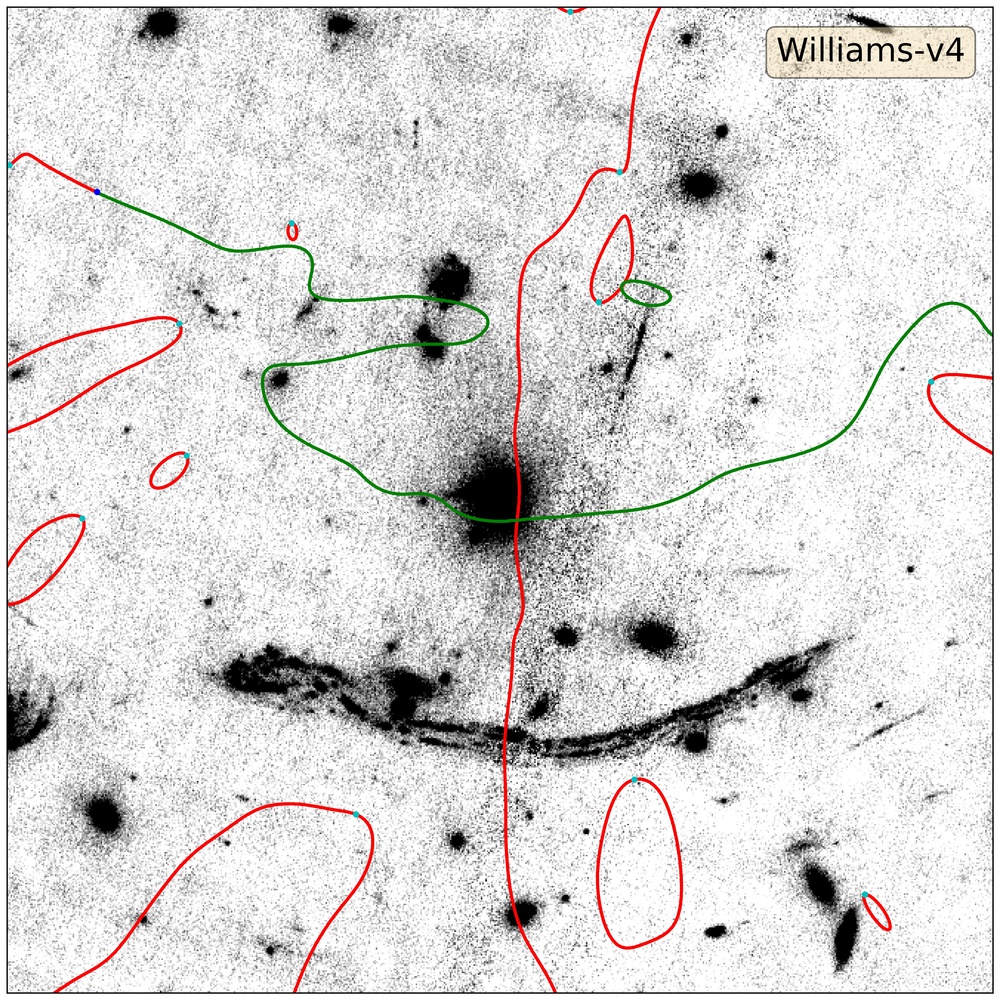}
  \includegraphics[width=\textwidth,height=8.5cm,width=8.5cm]{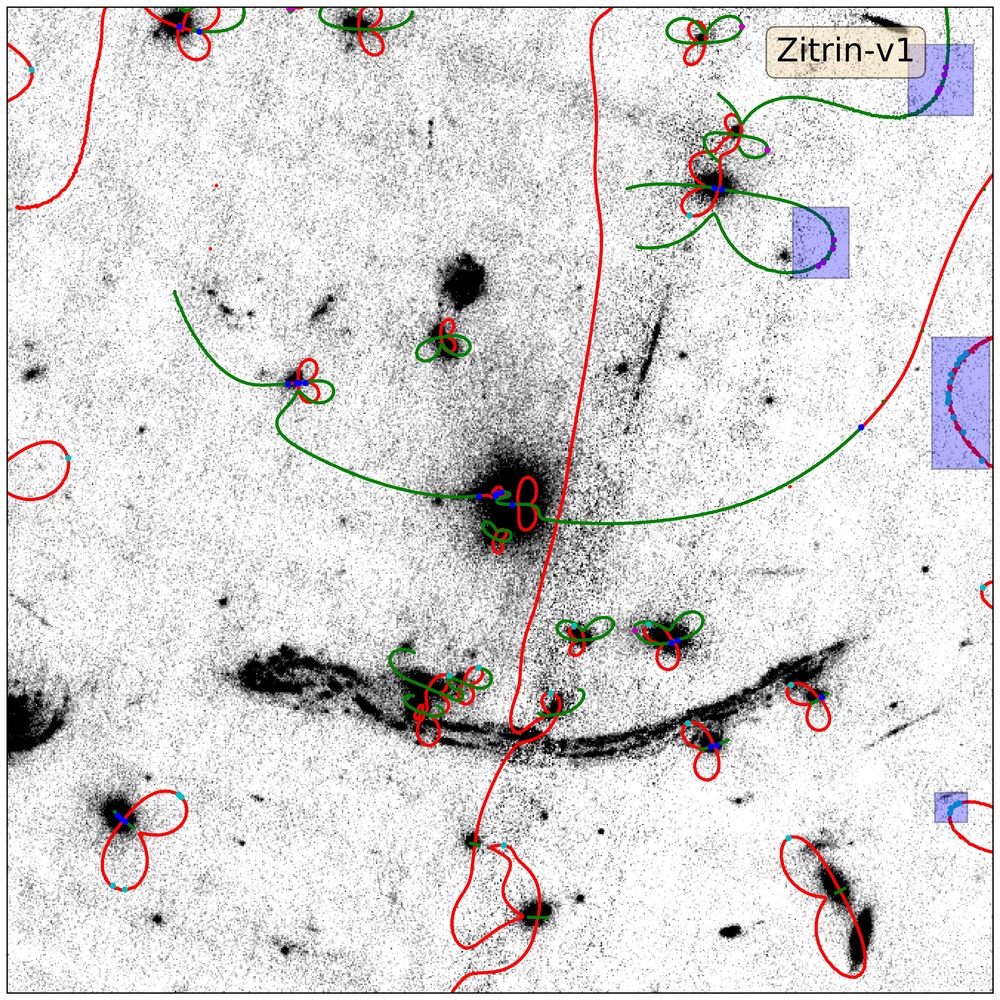}
  \caption{Singularity maps for the A370 cluster lens corresponding to mass 
  models provided by four different groups (Keeton, Sharon, Williams, Zitrin): 
  The red and green lines represent the $A_3$-lines corresponding to
  the tangential and radial cusps, respectively. The blue points
  denote the umbilics (hyperbolic and elliptic).  
  At hyperbolic umbilics, one red and one green line meet with each
  other, whereas   at elliptic umbilic, three red and three green
  lines meet. The cyan and magenta points represent the swallowtail
  singularities corresponding to the $A_3^\alpha$ and
  $A_3^\beta$-lines.  The shaded regions in the lower right panel mark
  the noisy region in the singularity map. These regions are not
  included in further calculations.
  In each panel, the background is the cluster image in the 
  F435W band.}
    \label{fig:a370 singularity map}
\end{figure*}

\section{Singularities in gravitational lensing}
\label{sec:singularities}

We briefly review different kinds of singularities that occur in
gravitational lensing in order to set up notions and notations.
For a detailed discussion, please see paper I, and for a pedagogic
discussion, you may see \citet{Schneider1992}.

The magnification factor of an image formed at $\mathbf{x}$ in lens
(image) plane is given by: 
\begin{equation}
\mu\left(\mathbf{x}\right)  =
\frac{1}{\left(1-a\alpha\right)\left(1-a\beta\right)}, 
\label{eq:(magnification)}
\end{equation}
where $a={D_{ds}}/{D_{s}}$ and $\alpha$ and $\beta$ ($\alpha \geq
\beta$) are the eigenvalues of the deformation tensor, $\psi_{ij}$ (a
$2\times2$ symmetric matrix made of second order partial derivatives
of the lens potential) and $D_{ds}$ and $D_s$ are the angular diameter
distances between lens and source, and observer and source,
respectively. 
The points in the lens plane where the magnification factor goes to 
infinity are the singular points of the lens mapping. 
These points form smooth closed curves known as critical curves  
in the lens plane. 
The corresponding curves in the source plane are known as caustics.
The caustics are also closed curves, though not necessarily smooth:
these are made up of smooth segments with cusps.
The smooth parts of caustics are called folds.  
 
As discussed in the paper I, there are two different kinds of
singularities in gravitational lensing: stable and unstable. 
Fold and cusp fall into the stable category as they are present for 
all possible source redshifts if the lens is critical.
The set of points corresponding to cusps in the lens plane form lines
known as $A_3$-lines.  
On the other hand, each of the unstable singularities (also known as
point singularities) only exist for specific source redshifts for a
given lens system.
All the point singularities are located on the $A_3$-lines. 
One can classify different point singularities using eigenvalues and
eigenvectors of the deformation tensor.
These different point singularities, along with the $A_3$-lines,
constitute a singularity map for a given lens model.
In lens plane, $A_3$-lines locate the points where the gradient of the 
deformation tensor eigenvalue is orthogonal to the corresponding eigenvector, 
i.e., $n_\lambda.\nabla_x\lambda=0$.
There are two $A_3$-lines, one corresponding to $\alpha$ and the other
corresponding to $\beta$ eigenvalue of the deformation tensor.
In the source plane, these two $A_3$-lines correspond to the cusps on
tangential and radial caustics, respectively. 

The point singularities not only satisfy the $A_3$-line condition but
also satisfy additional criteria depending on the type of the
singularity. 
For example, the swallowtail singularities indicate the points where
eigenvector $n_\lambda$ of the deformation tensor is tangent to the
corresponding $A_3$-line.  
The corresponding characteristic image formation is a tangentially or 
radially elongated arc made of four images (please see paper I for
details). 
On the other hand, the hyperbolic and elliptic umbilics denote the
point where $A_3$-lines corresponding to the different eigenvalues
meet with each other.  
At hyperbolic umbilic, two $A_3$-lines (one corresponding to $\alpha$
and one corresponding to $\beta$ eigenvalue) meet with each other,
whereas at elliptic umbilic six $A_3$-lines (three corresponding to
$\alpha$ and three corresponding to $\beta$ eigenvalue) meet with each
other.  
The characteristic image formation for hyperbolic umbilic is a ring
shaped (not Einstein ring) structure made of four images.
Whereas elliptic umbilic gives Y-shaped seven image configuration with
six radially elongated images with respect to the seventh central
image (please see paper I for details).  
As we know that $A_3$-lines trace the location of cusps in the source
plane, these point singularities represent the creation of an extra
pair of cusps or an exchange of cusps between tangential and radial
caustics in the source plane. 
These point singularities depend on the second and higher-order
derivatives of the lens potential, hence, they are very sensitive to
the lens potential. 
It is noteworthy that to date, we have only observed one
characteristic image formation near hyperbolic
umbilic~\citep{2008A&A...489...23L, 2009MNRAS.399....2O}, a handful of
image formations near swallowtail
singularity~\citep{1998MNRAS.294..734A, 2010A&A...524A..94S} and (to
the best of our knowledge) no image formation near elliptic umbilics.  

A singularity map is a compact representation of a given lens model,
and as discussed above, by locating $A_3$-lines and the point
singularities,  it finds all the high magnification regions in the
lens plane.  
These $A_3$-lines are the obvious targets for the deep-surveys for 
the given lens. 
In the following sections, we will construct singularity maps for 
different cluster lenses and look at the effects of the various mass 
reconstruction methods on the singularity map for a given lens.

\begin{figure*}
  \includegraphics[width=\textwidth,height=8.5cm,width=8.5cm]
  {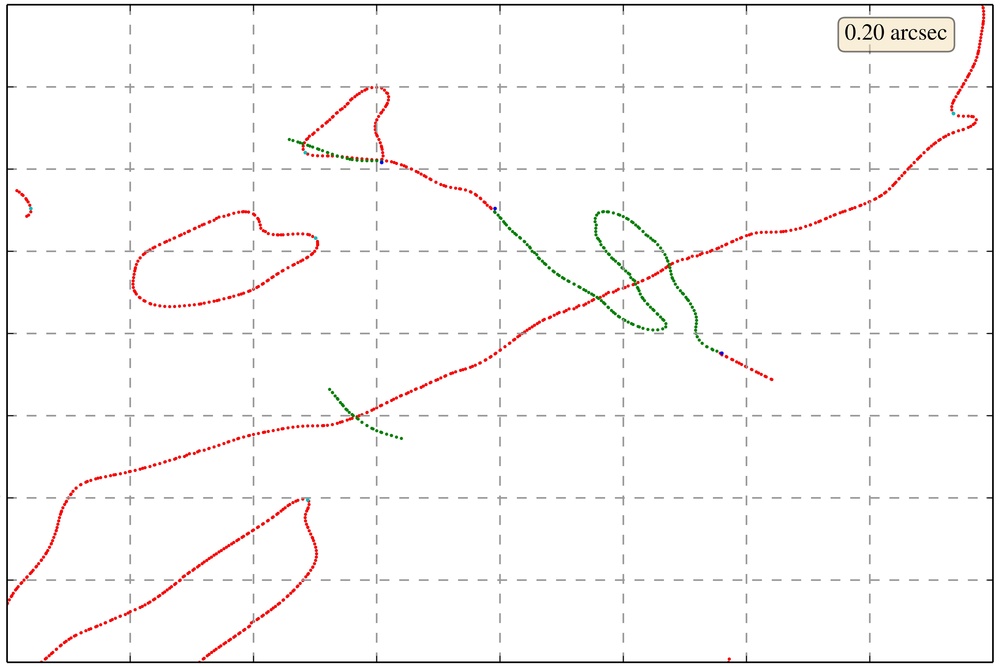}
  \includegraphics[width=\textwidth,height=8.5cm,width=8.5cm]
  {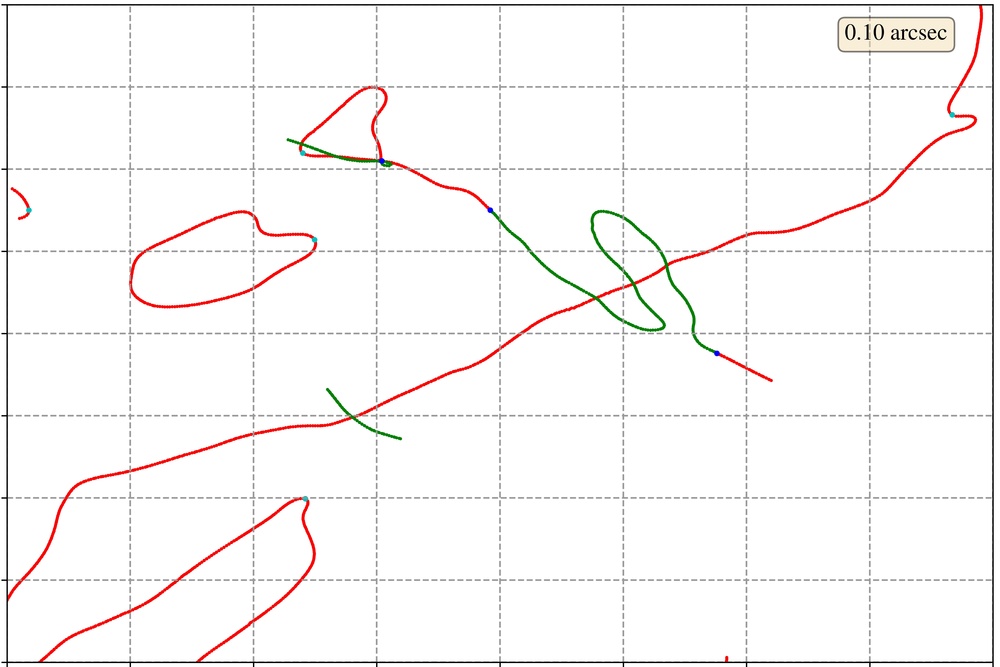}
  \includegraphics[width=\textwidth,height=8.5cm,width=8.5cm]
  {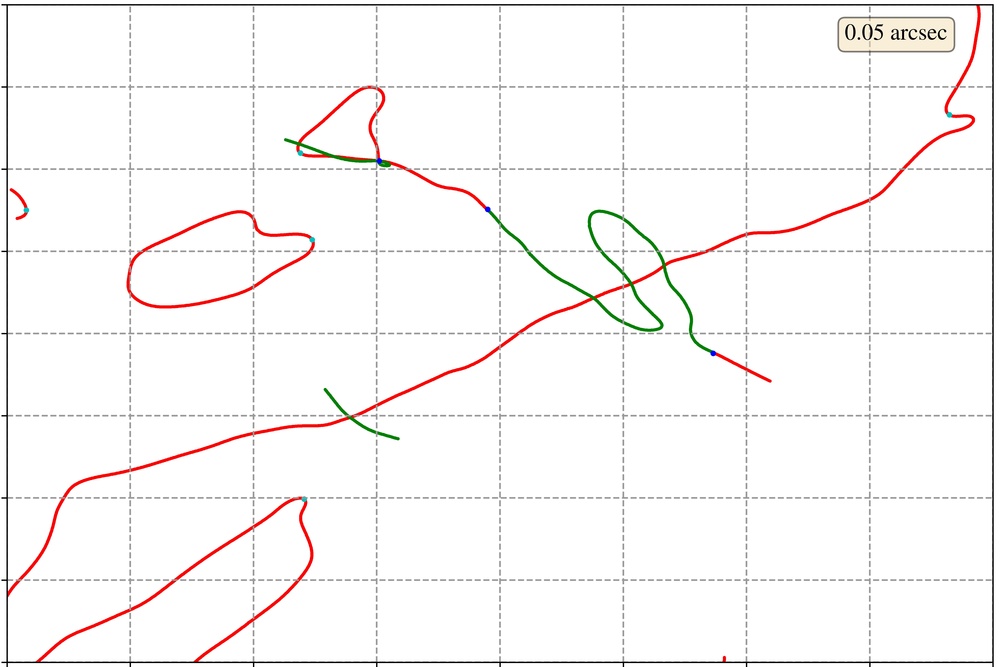}
  \includegraphics[width=\textwidth,height=8.5cm,width=8.5cm]
  {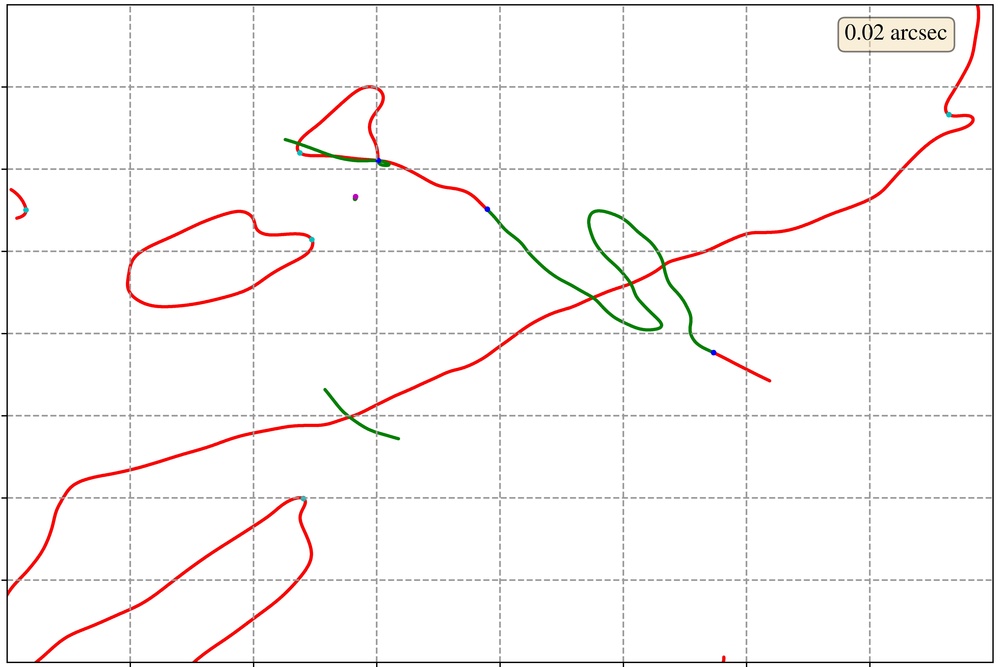}
  \caption{Singularity map for the MACS1149 cluster lens corresponding
    to four different resolution values,
    $0.20'',\:0.10'',\:0.05'',\:0.02''$. The color scheme is similar to the
    Figure~\ref{fig:a370 singularity map}.  As expected, increasing the
    resolution of the mass map helps in resolving the small scale
    structures in the singularity maps. Increase in resolution does
    not introduce any significant extra structures in the singularity
    map.} 
    \label{fig:map stability}
\end{figure*}

\begin{figure*}
  \includegraphics[width=\textwidth,height=7.cm,width=8.5cm]
  {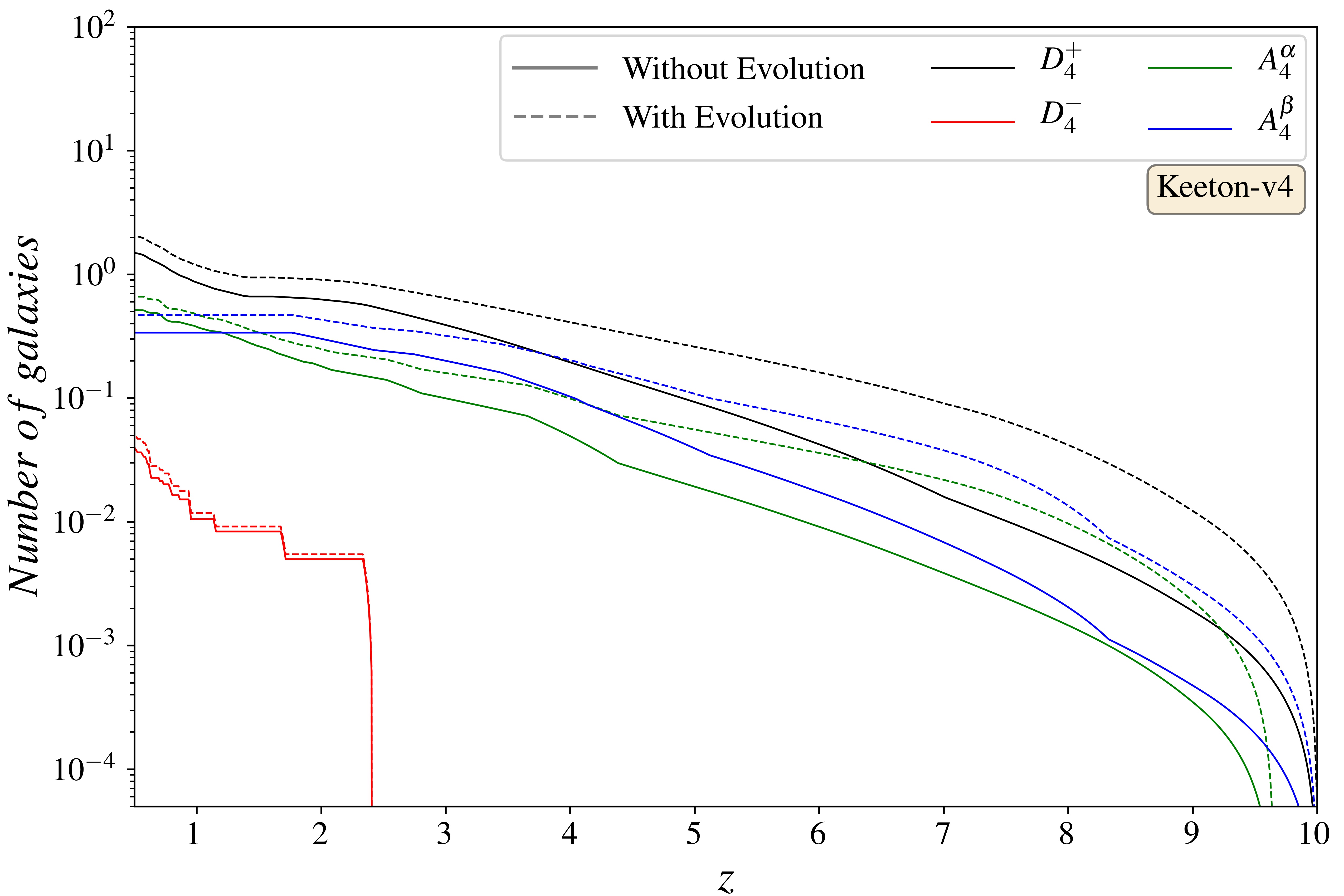}
  \includegraphics[width=\textwidth,height=7.cm,width=8.5cm]
  {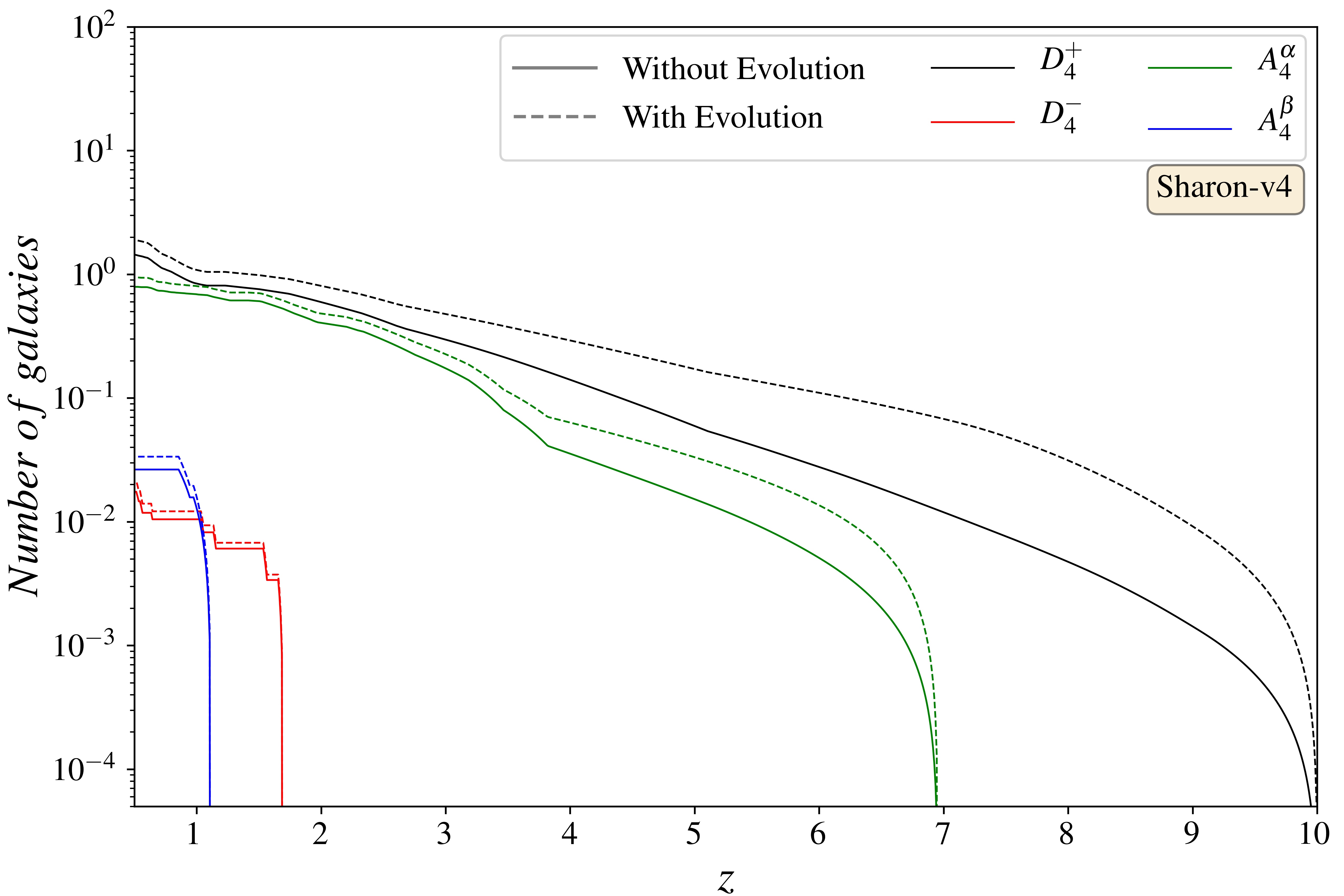}
  \includegraphics[width=\textwidth,height=7.cm,width=8.5cm]
  {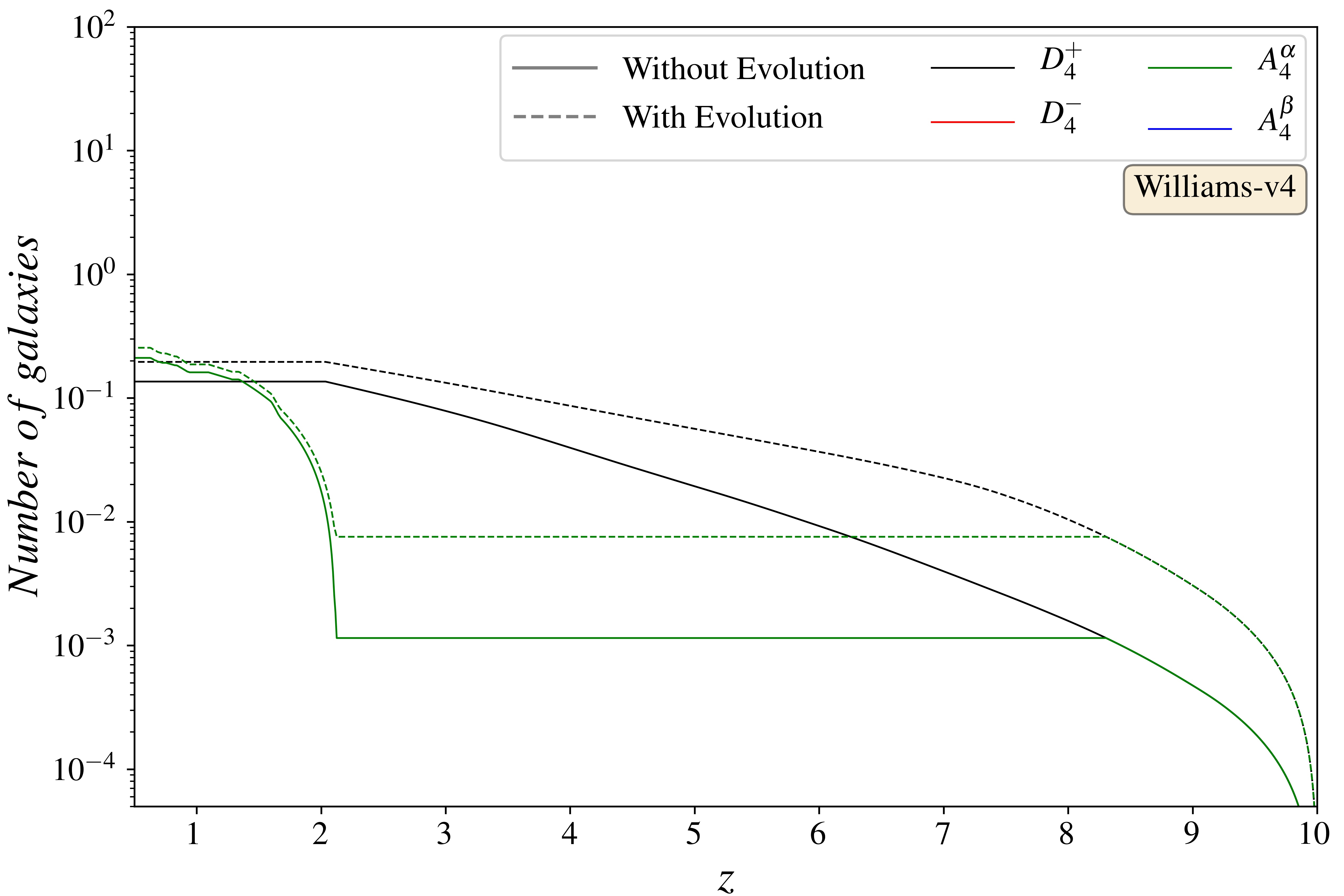}
  \includegraphics[width=\textwidth,height=7.cm,width=8.5cm]
  {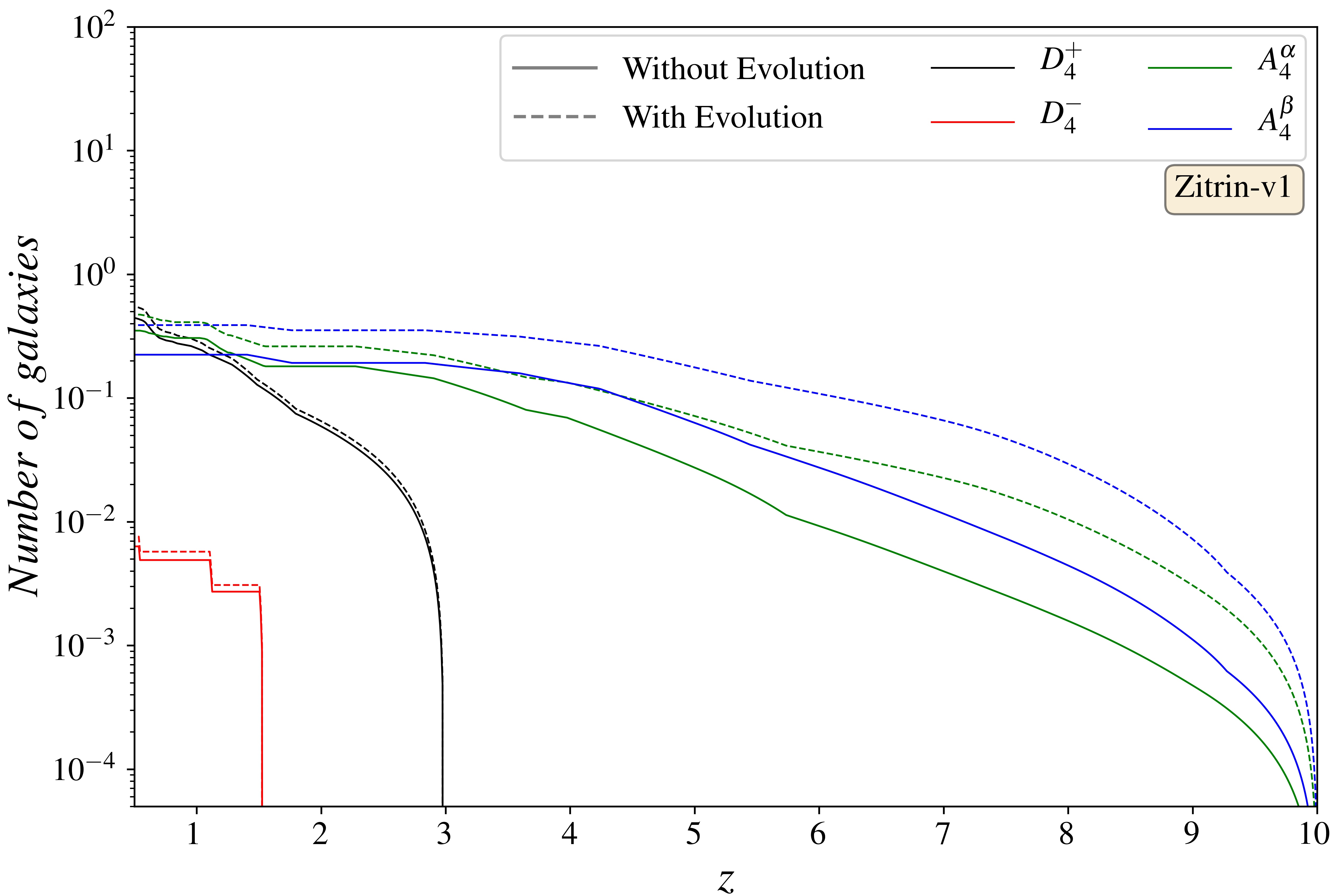}
  \caption{The cumulative number of source galaxies near point singularities
    as a function of redshift for A370 galaxy cluster: the $y$-axis shows
    the number at redshifts higher than $z$. 
  Different panels are 
  corresponding to different singularity maps in Figure~\ref{fig:a370 
  singularity map}, respectively. The solid lines represent the galaxy numbers
  calculated using the fiducial model used in C18, whereas the dashed lines 
  indicate the galaxy numbers calculated using the model with evolving 
  feedback (please see C18 for more details). The black and red lines denote the
  cumulative galaxy numbers corresponding to the hyperbolic and elliptic umbilic
  point singularities, respectively. Similarly, green and blue lines represent 
  the cumulative galaxy numbers corresponding to the swallowtail singularities 
  for $A_3^\alpha$- and $A_3^\beta$-lines, respectively.}
    \label{fig:galaxy numbers}
\end{figure*}

\begin{figure*}
  \includegraphics[width=\textwidth,height=7.0cm,width=8.5cm]
  {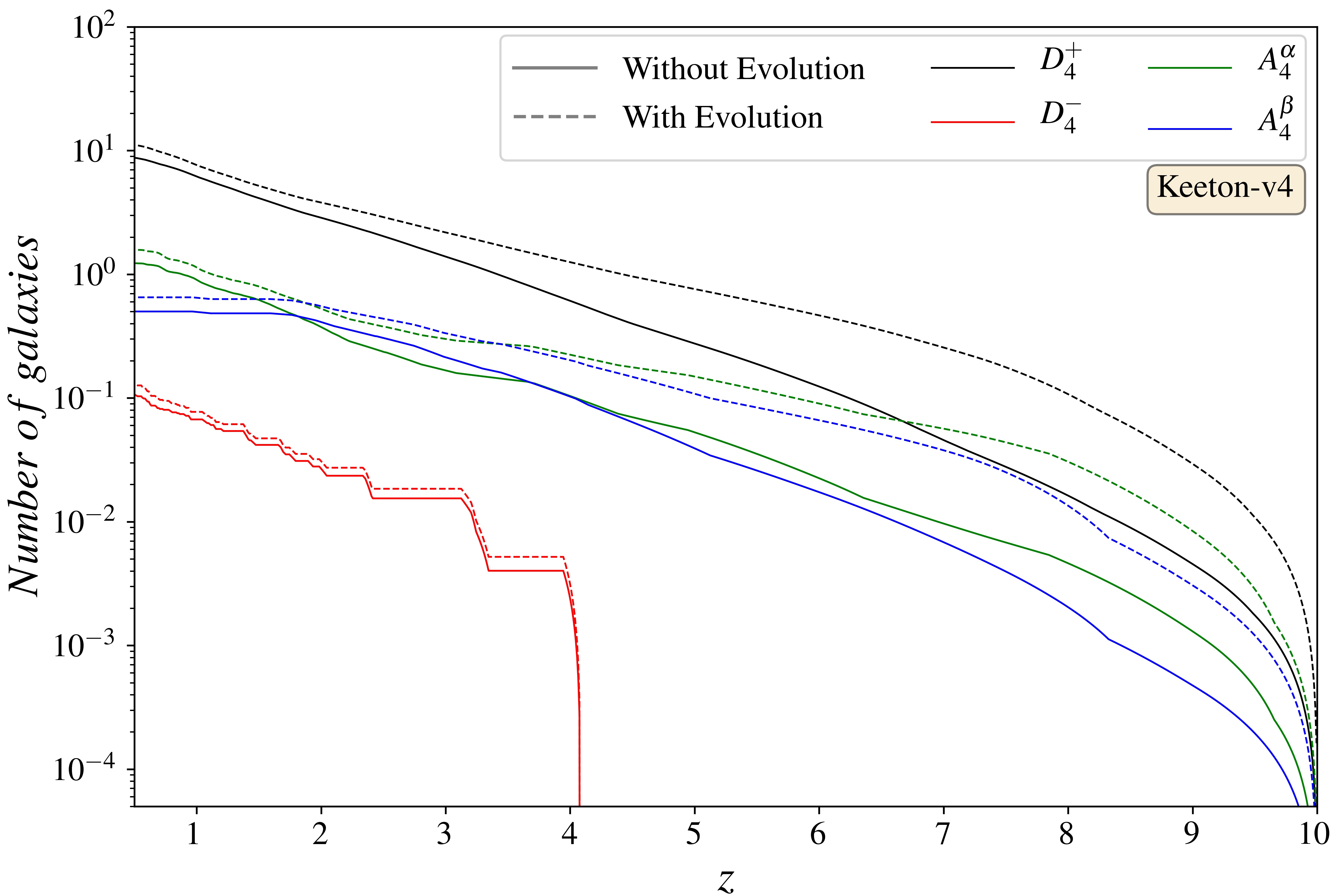}
  \includegraphics[width=\textwidth,height=7.0cm,width=8.5cm]
  {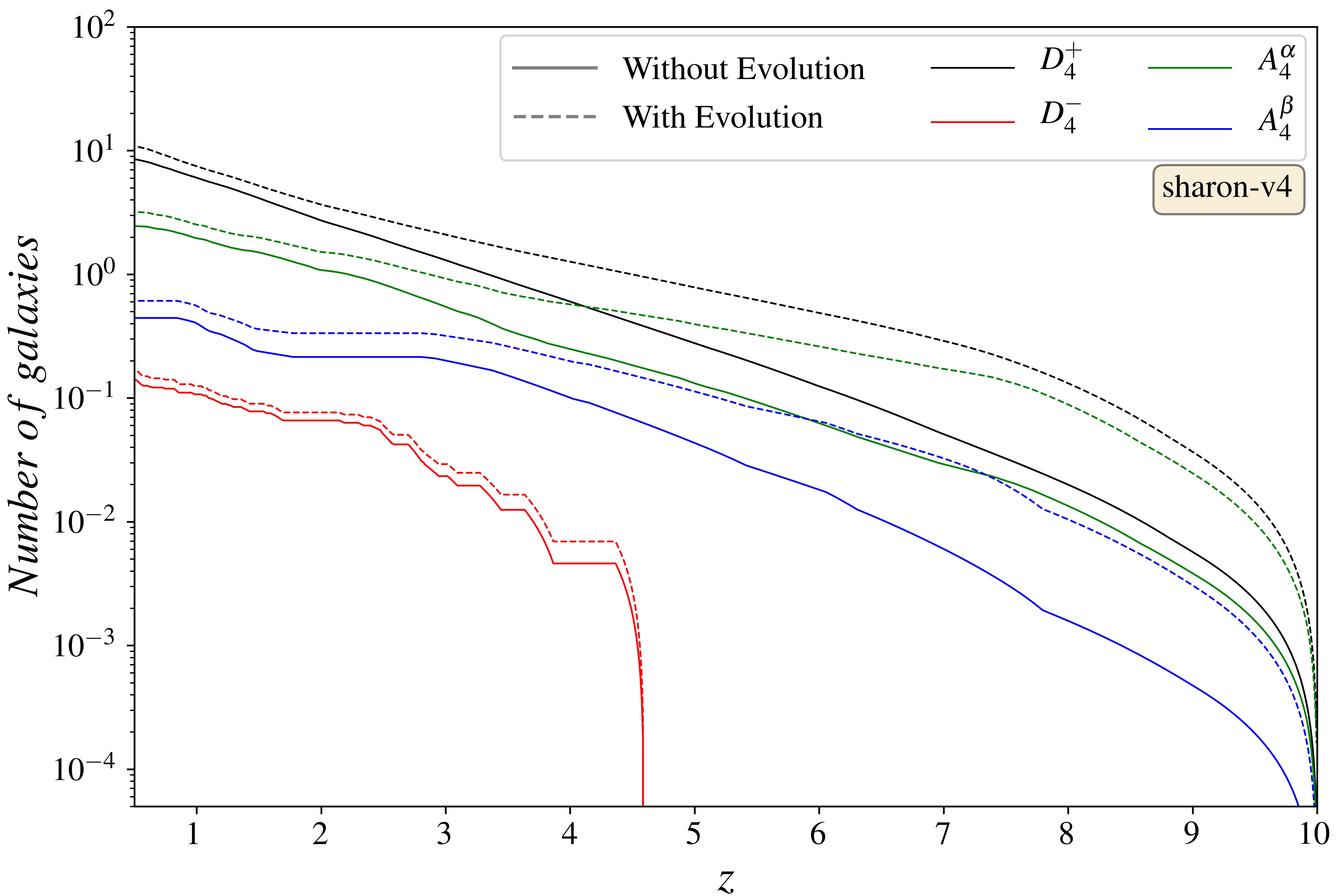}
  \includegraphics[width=\textwidth,height=7.0cm,width=8.5cm]
  {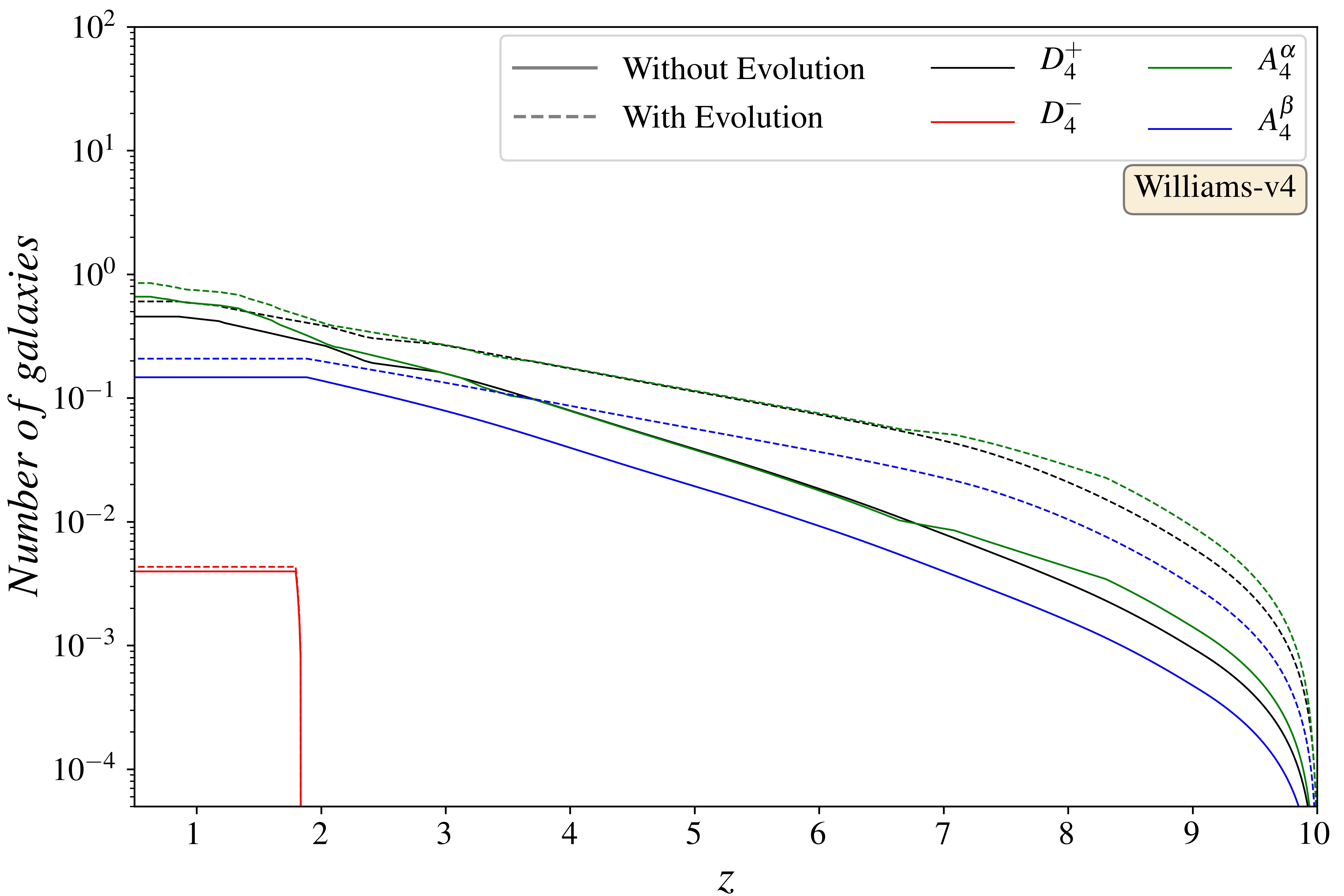}
  \caption{The total cumulative number of the source galaxies near point
   singularities as a function of redshift corresponding to the five of the 
   HFF clusters, namely, A370, A2744, AS1063, MACS0416, and MACS1149: the
   $y$-axis shows the number at redshifts higher than $z$. 
      The left, right, and bottom panels show the total galaxy numbers calculated 
   using the Keeton, Sharon, and Williams group mass models for each 
   cluster, respectively.
   Similar to Figure~\ref{fig:galaxy numbers}, the solid lines represent the 
   galaxy numbers calculated using the fiducial model used in C18, whereas the 
   dashed lines indicate the galaxy numbers calculated using the model with 
   evolving feedback. 
   The black and red lines denote the cumulative galaxy numbers corresponding 
   to the hyperbolic and elliptic umbilic point singularities, respectively. 
   The green and blue lines represent the cumulative galaxy numbers corresponding 
   to the swallowtail singularities for $A_3^\alpha$- and $A_3^\beta$-lines,
   respectively.}
    \label{fig:cumulative galaxy numbers}
\end{figure*}

\section{Cluster lenses}
\label{sec:clsuter lenses}

In this section, we briefly discuss the cluster lenses used to construct 
and study the singularity maps. 
Preliminary analysis in this direction consisting of ideal lens 
models and one real cluster lens, Abell 697, has been presented in 
paper~I. 
In the present analysis, we selected ten clusters for a detailed study 
of their singularity maps. 
Five out of these ten clusters were chosen from the \textit{Hubble 
Frontier Fields (HFF) survey} \citep{2017ApJ...837...97L}
\footnote{\url{https://archive.stsci.edu/prepds/frontier/}}
and the other five were chosen from the \textit{Reionization Lensing 
Cluster Survey (RELICS)} \citep{2019ApJ...884...85C}
\footnote{\url{https://archive.stsci.edu/prepds/relics/}}. 
The cluster lenses used in current study are described in
Table~\ref{tab:cluster_lenses}.   
The table provides relevant details, e.g,, the resolution and the
version of the mass models corresponding to various groups of
modelers. 

The HFF program targets a total of six massive merging clusters to
study the distant, faint sources and the cluster dynamics (please
see~\citet{2017ApJ...837...97L} for further details).  
Our analysis requires a model for the gravitational lenses.  
Mass reconstruction of the clusters has been attempted by multiple
groups~\citep{2005MNRAS.360..477D, 2011MNRAS.417..333M, 
2014MNRAS.443.1549J, 2014ApJ...797...48J, 2014MNRAS.443.3631M, 
2015ApJ...800...38G, 2015ApJ...799...12I, 2016ApJ...831..182H, 
2017A&A...600A..90C, 2018ApJ...855....4K} 
using different approaches.
The data from the observations is finite, one cannot model these
cluster lenses with arbitrary precision and resolution.
Different groups use different methods to reconstruct the cluster lens
mass distribution; for example, some groups use parametric modeling
(including the light distribution of cluster galaxies, some preferred
profile for the mass of cluster substructures) whereas some other
groups use the non-parametric approach which does not rely on any
assumption.
Some groups also use hybrid methods which take inputs from both
parametric and non-parametric approaches (please see
\citet{2017MNRAS.465.1030P, 2017MNRAS.472.3177M} for a comparison of
different modeling techniques). 
In our work, for five different HFF clusters, we used 
best-fit lens mass models provided by 
four different teams: Keeton, Sharon~\citep{2007NJPh....9..447J, 
2014ApJ...797...48J}, Williams~\citep{2007MNRAS.380.1729L} and 
Zitrin~\citep{2009MNRAS.396.1985Z, 2013ApJ...762L..30Z} (in the case of 
Zitrin group, we used mass models reconstructed using NFW profile, 
i.e., zitrin$\_$nfw) to construct the singularity maps for an HFF
cluster. 
We chose the central region of every cluster for our analysis, as these 
regions are responsible for the strong lensing. The size of the central 
part which we chose depends on the resolution of the lens model. 
In our analysis, we decided to use square regions with a side of $40''$. 
For low-resolution cluster lens models, the choice of a large area introduces 
noise, and this affects the reliability of that particular singularity 
map as the noise can introduce spurious singularities.
For the HFF cluster lenses, the central (ra, dec) values of the 
Sharon group are considered as standard values whenever we compare 
different singularity map for an HFF cluster. 
This is for ensuring uniformity and the choice does not arise from any
preference for one set of models.
As can be seen from Table~\ref{tab:cluster_lenses}, one of the HFF clusters,
MACSJ0717.5+3745, is not part of our current 
analysis as the corresponding singularity maps consist of a large number 
of spurious point singularities (please see section~\ref{ssec:singularity maps} 
for a discussion about noise in the singularity maps and its effects).

RELICS consists of a total of $41$ cluster lenses (please see section
2 in \cite{2019ApJ...884...85C}). 
For RELICS cluster lenses, at present, the mass modeling is done by three 
different groups using the Lenstool~\citep{2018ApJ...859..159C, 
2018ApJ...863..154P, 
2019ApJ...873...96M}, 
the LTM method~\citep{2018ApJ...858...42A, 
2019ApJ...874..132A,
2020ApJ...898....6A,
2018ApJ...863..145C}, 
and the Glafic tool~\citep{2020MNRAS.496.2591O}.
The Glafic mass models have a resolution of $0.1''$, whereas most of the 
Lenstool mass models have a resolution between $0.1''$ to $1.0''$. 
Therefore, for RELICS clusters, due to the low resolution, we cannot use the 
Glafic or the Lenstool mass models and cannot perform a comparison between 
different mass model reconstruction techniques (please see 
section~\ref{ssec:singularity maps} and~\ref{ssce:stability} for further 
details about the effect of mass model resolution). 
The Zitrin group mass models have a resolution of $0.06''$ but consist 
of a large number of spurious point singularities. 
The reason for the high number of spurious point singularities can be the 
that the higher-order derivatives of the lens potential are not well constrained at this resolution: note that all the other maps of these clusters use a coarser resolution.
Hence, in our current study, we only use one mass model for five
different RELICS clusters to construct the singularity maps 
(see~Table \ref{tab:cluster_lenses}).
The best-fit lens mass models considered here for RELICS 
clusters are parametric in nature and constructed by the Zitrin group using the
light-traces-mass (LTM) method with Gaussian smoothing
(zitrin$\_$ltm$\_$gauss). 
A more detailed comparison may become possible in future with the
availability of deep, high resolution images and construction of
corresponding lens models. 

\section{Results}
\label{sec:results}

Construction of singularity maps helps us in identifying the high
magnification regions in the lens plane of a given cluster lens, which
are obvious targets for the deep surveys. 
The high sensitivity of $A_3$-lines and point singularities to the
lens potential encouraged us to compare the singularity maps
corresponding to the different mass models for a cluster lens, which
are constructed using different techniques.  
Such a comparison provides us information about the effects of the 
reconstruction methods and the presence of substructures in a cluster 
lens on the singularity map.
Future observations of characteristic image formation around point
singularities may help distinguish between different models based on
present observations. 
In this section, we present results of our study of the construction
of singularity maps for the HFF and RELICS clusters, followed by a
comparison of different mass models for a cluster lens
(\S\ref{ssec:singularity maps}).  
In \S\ref{ssce:stability}, we study the stability of these
singularity maps against the mass model resolution and determined the
optimal resolution to construct the singularity map for a given lens. 
In \S\ref{ssec:cross section}, we estimate the number of source galaxies 
lying near the point singularities and the possibility of observing the 
corresponding characteristic image formation in upcoming all-sky
surveys followed by the constraints on the source redshift using the
point singularities in \S\ref{ssec: red_measure}.

\subsection{Singularity Maps}
\label{ssec:singularity maps}

We use the algorithm discussed in paper~I to construct the singularity 
maps for different lens models throughout this work.
In order to keep it concise, we only present singularity maps corresponding 
to one cluster lens (A370) in this article. 
The rest of the singularity maps are available online (along with all
the plots shown here) as supplementary material.  
Figure~\ref{fig:a370 singularity map}, represents the singularity maps for 
A370 corresponding to four different mass models (please see
Table~\ref{tab:cluster_lenses}).   
In every panel, the red and green line are the $A_3$-lines
corresponding to the $\alpha$ and $\beta$ eigenvalues of the
deformation tensor, respectively.  
The blue points show the hyperbolic and elliptic umbilics. 
The cyan and magenta colored points represent the swallowtail
singularities corresponding to the $\alpha$ and $\beta$ eigenvalues,
respectively. 
In each panel, the background is the cluster image in the F435W band.
Every map in Figure~\ref{fig:a370 singularity map} is a $40''\times40''$ 
central square region of the cluster with center coordinates (in degrees) 
$(39.971355,\, -1.582223)$. 
The source redshift ($z_s$) in the range between the lens redshift and
$z_s \le 10$ is used here.
Singularities in this range are shown in these plots. 
Hence, the $A_3$-lines trace the location of cusps for sources up to a 
redshift of ten.
 
As one can see from Figure~\ref{fig:a370 singularity map}, different 
singularity maps show differences in the $A_3$-line structures and the 
number of point singularities. 
However, one can still identify an overall $A_3$-line structure similar 
to an elliptical lens in every panel, which represents the entire cluster 
as an elliptical gravitational lens (please see paper~I for singularity map
corresponding to an elliptical lens). 
These differences arise mainly due to the fact that different groups use 
different mass reconstruction methods, and the number of substructures 
used by different teams is also different,
which is evident from Figure~\ref{fig:a370 singularity map}.
For example, mass models from Keeton, Sharon, and Zitrin groups are 
reconstructed using the parametric approach, which takes into account 
different properties of the cluster substructures as input and finds the 
best-fit parameters. 
On the other hand, the non-parametric reconstruction by Williams group 
uses no information regarding the cluster substructures as an input.
Hence, the singularity map corresponding to the Williams group shows 
the simplest $A_3$-line structure and least number of point singularities
as their reconstruction method does not give significant weightage to the 
presence of cluster substructures.
It is possible that substructure in their models is suppressed to some 
extent due to averaging over a large number of realizations in their
approach.
This is also evident from the fact that the best-fit mass model corresponding 
to the Williams group does not give a five-image configuration for a source 
at the giant arc's redshift ($z_s = 0.7251$) in A370. 
This can also be seen from the corresponding singularity map as there is no
swallowtail from the center to the giant arc, which can give rise to a 
five-image geometry.
On the other hand, all parametric mass models have swallowtails near the 
giant arc in the lens plane, which gives rise to five-image geometry.
Looking at the singularity maps corresponding to parametric mass models, 
one can see that different small scale structures introduce extra 
$A_3$-lines and point singularities in the singularity maps.

As mentioned above, due to the finite amount of observational data, 
one cannot achieve arbitrary high resolution during cluster lens 
mass reconstruction. 
The finite resolution of lens models also introduces a few problems 
in singularity maps.  
The first problem is the noise in the singularity map, which can be seen in 
the bottom right panel of Figure~\ref{fig:a370 singularity map}. 
The low resolution directly affects the shape of the $A_3$-lines, and it 
introduces spurious swallowtails point singularities as our algorithm 
first identifies the $A_3$-lines and uses these to locate swallowtail
singularities.  
To eliminate the effect of the noise, we do not include these spurious 
point singularities in our further calculations.
We mark these spurious point singularities utilizing the 
fact that in the absence of noise, the $A_3$-lines are smooth lines, 
which can be seen in every panel of Figure~\ref{fig:a370 singularity map} 
and~\ref{fig:map stability}. 
However, as the noise increases, points near the $A_3$-lines in the lens 
plane also contribute to the $A_3$-lines and affect the local shape of 
the $A_3$-lines. 
This, by definition, influences the number of swallowtails in the 
singularity map. 
We manually inspect every singularity map and mark such regions.
The other problem that has been introduced due to finite resolution are the 
missing point singularities. 
Sometimes when the distance between two similar kinds of point singularities 
is less than the grid size, our method is resolution limited and it
does not find these as two different point singularities. 
Instead, it only assigns one point singularity into that pixel. 
This mainly happens in the case of hyperbolic umbilics as pair of hyperbolic 
umbilic forms at the position of substructures in the singularity map. 
However, this does not affect the overall cross-section of hyperbolic 
umbilics significantly, as the number of such missed out points is tiny, 
and most of these points get critical at very low redshifts.
Hence, the contribution of these (left out) points in the cross-section 
is negligible.

The finite resolution of the mass models also affects the size of the 
singularity map. 
Hence, we are only able to construct singularity maps for the central region 
of the lens. 
As one goes away from the central region, the length of $A_3$-lines, and the 
total number of point singularities increases. 
However, the number of spurious structures introduced by noise also
increases.  
As a result, the number of point singularities that one can see in
Figure~\ref{fig:a370 singularity map}, should be considered as the
lower limit of the total number of point singularities that one
cluster lens has to offer.  

\subsection{Stability of Singularity Maps}
\label{ssce:stability}

One can deal with the above mentioned difficulties (noise and the left
out point singularities) by increasing the resolution of the mass
models.  
However, an increment in resolution can be computationally expensive. 
Apart from being computationally expensive, the other point that one
needs to take into account is the stability of the singularity maps.  
We know that the structures in a singularity map depend on the lens
potential and its higher-order derivatives in a non-linear fashion. 
Hence, the question arises, whether the increase in resolution can
introduce new structures in a singularity map?
Or do we reach convergence at some stage?
Addressing this question also helps us to find out the optimal
resolution for the construction of singularity maps. 
To answer this question, we constructed singularity maps corresponding
to mass models provided by the Williams group for the HFF clusters
with four different resolution values, $0.2'',\:0.1'',\:0.05'',$ and
$0.02''$. 
\footnote{The publicly 
available Williams group data files have a resolution of $\geq 0.2''$. 
However, using data files provided by Prof. Liliya Williams, we can
(in principle) resolve their mass models with an arbitrary resolution,
as their mass models are superposition of a large number of projected
Plummer density profiles.} 

Figure~\ref{fig:map stability}, shows the singularity map for MACS1149
with four different resolution values, $0.2'',\:0.1'',\:0.05'',$ and
$0.02''$.  
It is apparent that increasing the resolution of the mass models helps
us to better resolve the structures in the singularity maps.  
It does not introduce any new significant structures apart from the 
bottom right panel (resolution $0.02''$), where one extra loop of 
$A_3$-line corresponding to $\beta$-eigenvalue along with a
swallowtail makes an appearance. 
This is the case, at least in the case of non-parametric modeling.
This is because there is no structure smaller than the 
resolution $0.10''$ in Figure~\ref{fig:map stability}. 
However, in the case of parametric models, some structures are very
small even at a resolution of $0.05''$ and one can miss these
structures at a resolution of $0.10''$ or $0.20''$.  
This is also evident from the fact that in parametric models, even at a 
resolution of $0.05''$, some hyperbolic umbilics are missed. 
Hence, the optimal resolution to construct singularity maps, in the
case of both parametric and non-parametric modeling, should be of the
order of $0.02''$. 
This can be further confirmed in the case of parametric models by making
singularity maps with different resolutions (as we have done in
Figure~\ref{fig:map stability}) for non-parametric models.
For non-parametric mass models, somewhat low-resolution singularity
maps can also do the job. 
However, in the case of parametric modeling, one should construct mass
models with a resolution of at least $0.02''$ for construction of
singularity maps.
In general, the resolution of mass models should be better than or at
least equal to the resolution of observations used to arrive at the
map for completeness.

\subsection{Cross-Section}
\label{ssec:cross section}

Singularity maps can be used to study the variety of characteristic
image formation near point singularities.
This then becomes a template for searching different image types in
observations.
We expect the upcoming surveys to yield a number of systems, and a
quantitative prediction requires calculation of cross-section for each
type of singularity.

Once we draw the singularity maps for different mass models, the next
task is to determine the number of characteristic image formations
near different kinds of unstable singularities which can be observed
in surveys with different upcoming facilities. 
In order to make the estimate, we identify the range in redshift
around the critical redshift $z$ for each point singularity. 
We wish to choose the range such that the image formation can be
identified as characteristic of the given type of singularity. 
As discussed in the paper~I and mentioned above, image formation for
different point singularities evolves differently with redshift. 
Hence, the redshift interval in which one can observe the 
characteristic image formation will also be different for different 
point singularities.
Source redshift comes in the equation via distance
ratio, $a(=D_{s}/D_{ds})$.
Therefore we determine the corresponding distance ratio interval,
$\left[a-\delta a,a+\delta a\right]$ and use it to deduce the
appropriate redshift interval. 

In the case of hyperbolic umbilic, one can observe the characteristic
image formation at redshifts significantly smaller and larger than the
critical redshift, and we choose $\delta a = 0.1 a(z)$, where $a(z)$
is the distance ratio at the critical redshift.
The choice of $\delta a$ in this way also automatically takes into
account the fact that at small source redshifts, the caustics in the
source plane evolve more rapidly compared to the high redshifts.   
Hence, the distance ratio interval, $\left[a-\delta a,a+\delta
  a\right]$, for a point singularity is small for smaller source
redshifts.  
As we go towards higher source redshifts, the size of the distance ratio 
interval increases.
For swallowtail, one can only observe the characteristic image
formation beyond the critical redshift. 
So, in the case of swallowtail singularity, the distance ratio
interval modifies to $\left[a,a+\delta a\right]$, and the $\delta a$
is taken to be equal to $7\%$ of the distance ratio at the critical
redshift. 
For elliptic umbilic, the seven image Y-shaped image formation can
only be observed up to the critical redshift. 
Hence, for elliptic umbilic, the distance ratio interval modifies to
$\left[a-\delta a, a\right]$, and the $\delta a$ is equal to $0.5\%$
of the distance ratio at the critical redshift as elliptic umbilics
are highly sensitive to the redshift evolution.
 
Once we determine the redshift interval corresponding to the different
point singularities, we proceed to estimate the area in the source
plane around the caustics in which the source must lie to produce
characteristic image formation. 
As we know, the magnification factor for extended sources is
smaller than a point source~\citep{2019A&A...625A..84D}. 
Hence, if a compact source such as a star lies near the caustics in the 
source plane, then we get an observation of the characteristic 
image formation.
However, such sources are very rare in cluster
lensing~\citep{2018NatAs...2..334K}, and mostly we observe a galaxy as
a source. 
We are mainly considering galaxies as potential sources.
We consider a circular area in the source plane near the caustic
structure corresponding to the point singularities with a radius of
$5$~kpc.  

The above-chosen values of the distance ratio interval for different 
point singularities and source plane area near point singularities cannot 
be calculated mathematically. 
Hence, we manually estimate these numbers by observing a large number of 
lens systems (from galaxy to cluster scales) near point singularities and 
inspecting the corresponding image formations.
One should keep in mind that these numbers directly affect the calculated 
cross-section of point singularities and should be chosen very carefully.
Apart from that, as discussed in section~\ref{ssec:singularity maps} 
and~\ref{ssce:stability}, the finite resolution of the mass maps introduces 
the spurious point singularities, directly affecting the corresponding point
singularity cross-section. 
As one can see in supplementary material, the number of such spurious point
singularities is relatively high for RELICS clusters compared to the HFF
clusters.  
Hence, we do not calculate the point singularity cross-section for the RELICS 
clusters in our current work and restrict ourselves to the HFF clusters for point 
singularity cross-section estimation.

The remaining information that we need is the surface density of 
the observed galaxies as a function of redshift. 
However, the surface density of observed galaxies is sensitive to the
underlying models of galaxy formation and evolution as well as the
waveband and limiting magnitude.  
We consider a recent study by~\cite{2018MNRAS.474.2352C} (heareafter
C18) for JWST. 
In C18, a part of the work was to estimate the number of galaxies
observed in different bands of JWST, considering an exposure time
$10^4$ seconds (please see C18 for a detailed description). 
The useful quantity for our analysis, the surface density of observed
galaxies, is shown in Figures~9 and 10 of C18.
Here, for simplicity, we only consider one NIRCam filter, F200W, for
our analysis. 

Figure~\ref{fig:galaxy numbers}, represents the cumulative distribution
of the number of galaxies as a function of redshift 
which can provide us the characteristic image formation corresponding
to different point singularities for A370: we have plotted the numbers
expected at higher redshifts.
Different panels in Figure~\ref{fig:galaxy numbers} correspond to the 
singularity maps in Figure~\ref{fig:a370 singularity map} for
different mass models, respectively.
The solid lines represent the galaxy numbers with the fiducial model,
and the dashed lines correspond to a model with evolving feedback
(please see C18 for further details). 
The black, red, green, and blue lines are cumulative source galaxy
numbers that provide the characteristic image formations corresponding
to the hyperbolic umbilic, elliptic umbilic, swallowtail for
$A_3^\alpha$-line, and swallowtail for $A_3^\beta$-line,
respectively. 

As one can see from Figure~\ref{fig:galaxy numbers}, the probability
of finding a source galaxy at $z \geq 1$ with characteristic image
formation near hyperbolic umbilic or swallowtail for $A_3^\alpha$-line
is an order of magnitude higher for the Keeton/Sharon group mass
models compared to the Williams group mass model. 
This is due to the fact that the number of point singularities in the
Keeton/Sharon group mass models is much higher than the Williams group
mass models (please see Figure~\ref{fig:a370 singularity map}). 
On the other hand, the number of galaxies for the Zitrin group mass
model lies somewhat in between the galaxy numbers for Keeton/Sharon
and William models. 

This difference in the number of observed galaxy sources near point
singularities can also be seen in Figure~\ref{fig:cumulative galaxy
  numbers}.  
Figure~\ref{fig:cumulative galaxy numbers} represents the composite
cumulative distribution of the number of galaxies near point
singularities for five of the HFF clusters we are using for the
present study.
The left, right, and bottom panels correspond to the Keeton, Sharon, 
and Williams group mass models, respectively.
Here we did not calculate the galaxy numbers near point singularities
for the Zitrin group mass models (zitrin$\_$nfw) as the corresponding
singularity maps contain spurious point singularities.
However, the singularity maps are available online (except for
MACS1149 since the corresponding  (zitrin$\_$nfw) mass model is not
available). 
One can again see that the parametric mass reconstruction models give
an order of magnitude more source galaxies with characteristic image
formations at redshifts $\gtrsim\:1$ compared to the non-parametric
mass reconstruction models. 
From the Keeton and the Sharon group mass models, one expects to observe at 
least one image formation near swallowtail in all the HFF clusters and 
(on average) one image formation near hyperbolic umbilic in every HFF cluster. 
On the other 
hand, from the Williams group mass models, one expects to observe at least 
one image formation near swallowtail and one image formation near hyperbolic 
umbilic in all of the HFF clusters.
Given this pattern, singularity maps corresponding to the
non-parametric mass models can be used to estimate the minimum number
of expected characteristic image formation in the upcoming surveys. 

Our current analysis is based only on the best-fit models provided by 
different groups. 
Each of these models also has uncertainties associated with them due 
to the finite number of constraints 
available~\citep{2017ApJ...837...97L, 2017MNRAS.465.1030P}. 
These uncertainties affect the caustic structure in the source plane, 
and as a result, will also affect the point singularity cross-section.
However, we cannot estimate these uncertainties at present as we will 
need to construct the singularity map for each ensemble map to do that, 
and the corresponding potential or deflection maps are not available 
online (except for the Williams and Zitrin group).
A detailed investigation of the effects of these uncertainties is subject 
to our future study and will be presented in a forthcoming publication.

\subsection{Redshift Measurements}
\label{ssec: red_measure}

We have mentioned above that point singularities are critical only for
certain source redshifts, and the corresponding characteristic image
formation is only observable within a finite range for source
redshift.  
This encourages us to ask the question: can these point singularities be
used to constrain the source redshift as the corresponding
characteristic image formation is only visible within a specific
redshift range?  
In order to address this question, we consider the distance ratio
intervals, for characteristic image formation near different kinds of
point singularities. 
We find that point singularities constrain the source redshift
more strongly at smaller source redshifts than higher source redshifts.
For example, if a hyperbolic umbilic is critical at source redshift
one for a lens at redshift 0.35, then the characteristic image
formation is observable in the redshift range $\approx [0.85, 1.25]$. 
If the hyperbolic umbilic is critical at redshift below one, then the
redshift interval for the characteristic image formation further
narrows down considerably. 
On the other hand, if the hyperbolic umbilic is critical at redshift
five, then the image formation can be observed in the redshift range
$\approx [2.2, 10]$. 
The same argument can be also used for other point singularities.
However, for other point singularities the redshift range is smaller
than the redshift range for hyperbolic umbilic (please see 
subsection~\ref{ssec:cross section}) and these can provide stronger 
constraints on source
redshift than the hyperbolic umbilics, if observed.  
Hence, point singularities are more useful in constraining the source
redshift at smaller redshifts. 
One can also understand such behavior from the fact that the caustic
structure evolves more rapidly at smaller redshifts compared to higher
redshifts. 
The elliptic umbilic can be useful at higher redshifts; however, the
observational cross-section for the corresponding image formation is
negligible (please see Figure~\ref{fig:galaxy numbers}
and~\ref{fig:cumulative galaxy numbers}).
It is also important to keep the context of a known lens map for this
discussion, if there are uncertainties in the lens map then it may be
better to find out the source redshift to constrain the lens map.

\section{Conclusion}
\label{sec:conclusion}

We have constructed singularity maps corresponding to ten different
galaxy clusters selected from the HFF and the RELICS survey.
To construct singularity maps, we followed the algorithm developed and
discussed in the paper I. 
Such a singularity map traces all the optimal sites for the upcoming
deep surveys in the cluster lens plane as well as mark the locations
of all the point singularities. 
Point singularities are very sensitive to the lens potential as these
have a non-linear dependence on higher-order derivatives of the lens
potential. 
Hence, these are also sensitive to the mass reconstruction methods as
different methods use a different set of assumptions to construct mass
models. 
We have constructed singularity maps corresponding to five of the HFF
clusters (Table~\ref{tab:cluster_lenses}), considering mass models
reconstructed using both parametric (Keeton, Sharon, and Zitrin
groups) and non-parametric (Williams group) techniques.  
On the other hand, for five of the RELICS clusters, we only considered
one mass model (provided by the Zitrin group) for each cluster lens.
We find that the number of point singularities corresponding to
parametric and non-parametric mass models is very different.  
The parametric models give a large number of point singularities
compared to non-parametric models where only a handful of point
singularities are present (Figure~\ref{fig:a370 singularity map}).  
This also affects the estimated number of galaxy sources with
characteristic image formation as the parametric models yield an order
of magnitude large number of such sources compared to non-parametric
models (Figure~\ref{fig:galaxy numbers} and~\ref{fig:cumulative galaxy
  numbers}).
We suspect that the assumption of mass associated with each of the
galaxies in the lensing cluster is the reason for 
this~\citep{2007NJPh....9..447J}.
As the number of point singularities corresponding to the
non-parametric mass models is the least, one can use these models to
compute the lower limit on the observation of characteristic image
formation in the upcoming all-sky surveys.
We find that the number of point singularities is significantly higher
than estimates in earlier studies~\citep{2009MNRAS.399....2O}. 
Recently~\citet{2020arXiv200904471M} pointed out an order of magnitude 
discrepancy in substructure lenses in the observed galaxy clusters
(which are modeled using the parametric approach) and simulated galaxy clusters.
While the discrepancy with LCDM is disputed \citep{2021arXiv210112112B, 2021arXiv210112067R},
one can expect a significant difference in 
the number of point singularities that may arise as lens models are 
better constrained with more images.  

The key takeaway from our analysis is that the predicted number of
instances of point singularities in cluster lenses is likely to be much
higher than estimated earlier \citep{2009MNRAS.399....2O}. 
We expect to get at least one hyperbolic umbilic and one 
swallowtail image formation for a source at z > 1 for every five 
clusters with JWST. 
This estimation is based on the non-parametric mass models corresponding 
to the Williams group, and it can be considered as a lower limit since the 
number of point singularities is much higher in parametric models than 
non-parametric models. 

We have not considered galaxy lenses in our analysis. 
Each galaxy scale lens has at least one pair of hyperbolic umbilics
(as these are modeled as elliptical mass distributions with more
details thrown in) and sometimes a few swallowtails (if a substructure
also exists).  
Including the galaxy lenses will further increase the possibility of
observing the image formations near point singularities. 

Addition of substructure always adds more singular points, hence the
numbers and distribution of singular points can be connected with the
amount of substructure in clusters.
Image formation near a point singularity consists of multiple images
lying very close to each other in the lens plane: the characteristic
image formation being different for each type of singularity. 
This opens up the possibility of measuring the relative time delay 
between these images. 
Such measurements are possible even if the multiple images are not
well resolved~\citep{2008MNRAS.389..364B}. 
Along with the time delay analysis, one can also construct an atlas of
realistic image configurations near point singularities for training
and identification using machine learning programs in the upcoming
surveys~\citep{2019MNRAS.487.5263D}.  
These possibilities are the subject of our future studies and the
results will be presented in forthcoming publications.

\section{Acknowledgements}

AKM would like to thank Council of Scientific $\&$ Industrial Research 
(CSIR) for financial support through research fellowship  No. 524007. 
Authors are very grateful to Liliya Williams for providing the mass 
models for HFF clusters and their help.
Authors thank Adi Zitrin for answering queries related to mass models 
provided by their group.
Authors thank Prasenjit Saha for their useful comments on the manuscript.
Authors thank the anonymous referee for useful comments.
We acknowledge the HPC@IISERM, used for some of the computations 
presented here.
This research has made use of NASA's Astrophysics  Data  System 
Bibliographic Services.
This work utilizes gravitational lensing models produced by 
PIs Bradac, Natarajan $\&$ Kneib (CATS), Merten $\&$ 
Zitrin, Sharon, Williams, Keeton, Bernstein and Diego, and the GLAFIC group. 
This lens modeling was partially funded by the HST Frontier Fields program 
conducted by STScI. STScI is operated by the Association of Universities 
for Research in Astronomy, Inc. under NASA contract NAS 5-26555. 
The lens models were obtained from the Mikulski Archive for Space 
Telescopes (MAST).
This work is based on observations taken by the RELICS Treasury 
Program (GO 14096) with the NASA/ESA HST, which is operated by 
the Association of Universities for Research in Astronomy, Inc., 
under NASA contract NAS5-26555.

\section{Data Availability}

The clusters mass models used in this article are available at the official 
site of \textit{Hubble Frontier Fields (HFF) survey} 
\footnote{\url{https://archive.stsci.edu/prepds/frontier/}}
and  \textit{Reionization Lensing Cluster Survey (RELICS)}
\footnote{\url{https://archive.stsci.edu/prepds/relics/}}. 
The high resolution mass models for the Williams group is available 
on request from the modelers.

\bsp	
\label{lastpage}
\end{document}